\documentclass[twocolumn,showpacs,preprintnumbers,amsmath,amssymb]{revtex4}

\usepackage{graphicx}
\usepackage{dcolumn}
\usepackage{bm}

\begin{document}
\title{\bf{Unambiguous discrimination between mixed quantum states based on \\ programmable quantum state discriminators}}
\author{Hongfeng Gan and Daowen Qiu}\email{issqdw@mail.sysu.edu.cn  (D.
Qiu).}
 \affiliation{%
 Department of
Computer Science, Zhongshan University, Guangzhou 510275,
 People's Republic of China
}%
\date{\today}

\begin{abstract}
 We discuss the problem of designing an unambiguous programmable
 discriminator for mixed quantum states. We prove that there does not exist such a universal unambiguous programmable
 discriminator for mixed quantum states that has two program registers and one data register. However, we find that we can use the idea of programmable
 discriminator to unambiguously discriminate mixed quantum states.
The research shows that by using such an idea, when the success
probability for discrimination reaches  the upper bound,
 the success probability is better than what we do not use the idea to do, except for some special cases.
  \end{abstract}
 \maketitle
  \section{\label{Introduction}Introduction}

The discrimination of quantum states is a basic task in quantum
information and quantum communication [1]. A great deal of
attention has been attracted into this field this years,
especially the {\it unambiguous discrimination} (UD) of quantum
stats. UD is a sort of discrimination that never gives an
erroneous result, but sometimes it may fail. In the case of pure
states, UD has been widely considered. In the case of two pure
states, the optimum measurement for the UD of two pure states was
found decade ago [2-5]. A sufficient and necessary condition for
unambiguously distinguishing arbitrary pure states and upper bound
on the success probability for UD of arbitrary pure states have
also been given (see, for example, [6] and some related references
therein). Indeed, a complete overview of UD of pure states can be
found in two recent review articles [7]. In the case of mixed
quantum states, lots of work also has been done this years [8-14],
which focuses on the upper bound and how to get the upper bound of
the success probability for discrimination. For the case of two
mixed quantum states, a necessary and sufficient condition for
discriminating two mixed states to reach upper bound has been
derived in  [12].

As we know, if we want to unambiguously discriminate quantum
states, we need construct some positive operator valued
measurements (POVMs) according to the states. However, if the
states are unknown, we can not construct such POVMs, which means
that we can not discriminate unknown states directly. Recently, a
programmable quantum state discriminator for unambiguous
discrimination was first proposed by Bergous and Hillery [15] to
resolve this problem. Bergous and Hillery's discriminator is a
fixed measurement that has two program registers and one data
register. The quantum states in the data register is what we want
to identify, which is confirmed to be one of the two states in
program registers. That is to say, if we want to discriminate two
states $|\psi_1\rangle$ and $|\psi_2\rangle$, we assign the two
sates into the two program registers, and the data register is
assigned with the state which we want to identify. Here we have no
idea of these two states. Now we have two input states
\begin{align}
\begin{split}
&|\psi^{in}_1\rangle=|\psi_1\rangle|\psi_2\rangle|\psi_1\rangle,
|\psi^{in}_2\rangle=|\psi_1\rangle|\psi_2\rangle|\psi_2\rangle.
 \end{split}
 \end{align}
It is easy to see that if we can discriminate
$|\psi^{in}_1\rangle$ and $|\psi^{in}_2\rangle$, then we can
discriminate states $|\psi_1\rangle$ and $|\psi_2\rangle$. Bergous
and Hillery's discriminator makes this target successful with a
fixed measurement.

Based on Bergous and Hillery's discriminator, Zhang {\it et al}
[16] recently presented an unambiguous programmable discriminator
for $n$ arbitrary quantum states in an $m$-dimensional Hilbert
space, where $m\geq n$. If $m=n$, an optimal unambiguous
programmable discriminator for $n$ arbitrary states was given in
[16]. Notably, the unambiguous programmable discriminator for two
states with a certain number of copies has been discussed in
[17,18].

However, all the discriminators mentioned above concentrate on
pure states. As we are aware, the unambiguous programmable
discriminators for mixed quantum states still have not been
discussed. In this paper, we try to deal with the problem of
designing an unambiguous programmable discriminator for mixed
quantum states. Our purpose is to see whether or not an
programmable unambiguous discriminator for mixed quantum states
can be realized.

This paper is organized as follows: In Sec. \ref{Protocols}, we
prove that there does not exist an unambiguous programmable
discriminator for mixed quantum states that has two program
registers and one data register. Then, however, in Sec.
\ref{Protocols1}, we find that we can still use the idea of
programmable quantum state discriminator to unambiguously
discriminate mixed quantum states. The research shows that by
using this idea, when the success probability for discrimination
reaches the upper bound, the success probability is better than
what we do not use the idea to do, except for some special cases.
At last, we conclude the paper with a short summary.

 \section{\label{Protocols} Nonexistence of programmable discriminator for mixed states based on Bergous and Hillery's model}
First we try to design an unambiguous programmable discriminator
for mixed quantum states based on Bergous and Hillery's model
[15]. Our purpose is to see whether  or not such an unambiguous
programmable discriminator for mixed quantum states can be
realized. To begin with, we prove a theorem here.

\textit{Theorem 1.} Two mixed quantum states $\rho_1$, $\rho_2$
can be unambiguously discriminated if and only if $\rho^{in}_1$,
$\rho^{in}_2$ can be unambiguously discriminated, where
\begin{align}
\begin{split}
&\rho^{in}_1=\rho_1\otimes \rho_2\otimes \rho_1,\hskip 2mm
\rho^{in}_2=\rho_1\otimes \rho_2\otimes \rho_2.
 \end{split}
 \end{align}

\textit{Proof.} First let
\begin{align}
\begin{split}
&\rho_1=\sum_{i=1}^{n_1}\alpha_i|\varphi_i\rangle\langle\varphi_i|,\hskip
2mm \rho_2=\sum_{j=1}^{n_2}\beta_j|\psi_j\rangle\langle\psi_j|
\end{split}
 \end{align}
be the spectral decompositions of $\rho_1$, $\rho_2$. Then
\begin{align}
\begin{split}
&\rho^{in}_1=\sum_{i,j,k}\alpha_i\beta_j\alpha_k|\varphi_i\psi_j\varphi_k\rangle\langle\varphi_i\psi_j\varphi_k|,\\
&\rho^{in}_2=\sum_{i,j,l}\alpha_i\beta_j\beta_l|\varphi_i\psi_j\psi_l\rangle\langle\varphi_i\psi_j\psi_l|.
\end{split}
 \end{align}
where $i,k=1,...,n_1$, $j,l=1,...,n_2$. Clearly,  formula (4) is
also the spectral decompositions of $\rho^{in}_1$ and
$\rho^{in}_2$.

Suppose that $\rho_1$, $\rho_2$ can be unambiguously
discriminated. Then there exist POVM elements $\Pi_0$, $\Pi_1$,
$\Pi_2$ such that $\Pi_0+$$\Pi_1+$$\Pi_2=I$ and
Tr$(\Pi_i\rho_j)=p_i\delta_{ij}$ for some $p_i>0$, where
$i,j=1,2$. Now we construct a new set of POVM elements
$\Pi^{in}_0=I^{'}\otimes\Pi_0$, $\Pi^{in}_1=I^{'}\otimes\Pi_1$,
$\Pi^{in}_2=I^{'}\otimes\Pi_2$, where $I^{'}$ denotes the identity
operator on $\rho_1\otimes \rho_2$. We can easily prove that
$\Pi^{in}_0+$$\Pi^{in}_1+$$\Pi^{in}_2=I$ and
Tr$(\Pi^{in}_i\rho^{in}_j)=p_i\delta_{ij}$ for the above $p_i>0$,
where $i,j=1,2$. It means that there exists a set of POVM elements
which can unambiguously discriminate $\rho^{in}_1$, $\rho^{in}_2$,
i.e., $\rho^{in}_1$, $\rho^{in}_2$ can be unambiguously
discriminated.

On the other side, suppose that $\rho^{in}_1,\rho^{in}_2$ can be
unambiguously discriminated. Then $supp(\rho^{in}_1)\neq
supp(\rho^{in}_1,\rho^{in}_2)$ and $supp(\rho^{in}_2)\neq
supp(\rho^{in}_1,\rho^{in}_2)$ [10]. Here
$supp(\rho_1,...,\rho_n)$ is defined by the Hilbert space spanned
by the eigenvectors of the mixed states $\rho_1,...,\rho_n$ with
corresponding nonzero eigenvalues. For $supp(\rho^{in}_2)\neq
supp(\rho^{in}_1,\rho^{in}_2)$, it means that there exist some
$i,j,k$, where $1\leq i,k\leq n_1$ and $1\leq j \leq n_2$,
satisfying
\begin{align}
|\varphi_i\psi_j\varphi_k\rangle\neq\sum_{i^{'},j^{'},l^{'}}a_{i^{'},j^{'},l^{'}}|\varphi_{i^{'}}\psi_{j^{'}}\psi_{l^{'}}\rangle
 \end{align}
where $i^{'}=1,...,n_1$, $j^{'},l^{'}=1,...,n_2$. Specifically, if
we choose $i^{'}=i$, $j^{'}=j$, then
\begin{align}
|\varphi_i\psi_j\varphi_k\rangle\neq|\varphi_i\psi_j\rangle\sum_{l^{'}}a^{'}_{l^{'}}|\psi_{l^{'}}\rangle,
 \end{align}
and, as a result,
\begin{align}
|\varphi_k\rangle\neq|\sum_{l^{'}}a^{'}_{l^{'}}|\psi_l^{'}\rangle.
\end{align}
It implies $supp(\rho_2)\neq supp(\rho_1,\rho_2)$. With similar
discussion, we can also have $supp(\rho_1)\neq
supp(\rho_1,\rho_2)$. Therefore, $\rho_1$,$\rho_2$ can be
unambiguously discriminated. This completes the proof.

In terms of \textit{Theorem 1}, we discuss whether or not there
exists an unambiguous programmable discriminator for mixed quantum
states based on Bergous and Hillery's model [15]. Indeed, we have
the following result.

\textit{Theorem 2.} There does not exist an unambiguous
programmable discriminator for mixed quantum states that has two
program registers and one data register.

\textit{Proof.} Suppose that there exists such an unambiguous
programmable  discriminator for mixed quantum states. Then there
also exists a fixed measurement that can unambiguously
discriminate $\rho^{in}_1,\hskip 1mm \rho^{in}_2$, where
$\rho^{in}_1=\rho_1\otimes \rho_2\otimes \rho_1,
\rho^{in}_2=\rho_1\otimes \rho_2\otimes \rho_2$, and
$\rho_1,\rho_2$ are guaranteed to be unambiguously discriminated.
We here assume that the fixed POVM elements are
$\Pi_0$,$\Pi_1$,$\Pi_2$, which satisfy
\begin{align}
\begin{split}
&\Pi_1\rho^{in}_2=0,\Pi_2\rho^{in}_1=0,\\
&Tr(\Pi_1\rho^{in}_1)>0,Tr(\Pi_2\rho^{in}_2)>0,\\
&\Pi_0+\Pi_1+\Pi_2=I,
\end{split}
 \end{align}
for any $\rho_1,\rho_2$ when they can be unambiguously
discriminated.

Now, we have three special mixed quantum states as follows
\begin{align}
\begin{split}
&\rho^{'}_1=a_1|\gamma_1\rangle\langle\gamma_1|+a_2|\gamma_2\rangle\langle\gamma_2|,\\
&\rho^{'}_2=b_1|\gamma_2\rangle\langle\gamma_2|+b_2|\gamma_3\rangle\langle\gamma_3|,\\
&\rho^{'}_3=c_1|\gamma_1\rangle\langle\gamma_1|+c_2|\gamma_3\rangle\langle\gamma_3|,
\end{split}
 \end{align}
where $\rho^{'}_1$, $\rho^{'}_2$, $\rho^{'}_3$ are mixed quantum
states in $m$-dimension Hilbert space ($m\geq3$), and
$\{|\gamma_1\rangle,|\gamma_2\rangle,|\gamma_3\rangle\}$ consists
of an orthonormal basis in this space. It is no doubt that any two
of these three stats can be unambiguously discriminated. Now we
use the discriminator to discriminate any two of these states.

(1) Let $\rho_1=\rho^{'}_1,\rho_2=\rho^{'}_2$. Then
$\rho^{in}_1=\rho^{'}_1\otimes \rho^{'}_2\otimes \rho^{'}_1,
\rho^{in}_2=\rho^{'}_1\otimes \rho^{'}_2\otimes \rho^{'}_2$.
According to (8), $\Pi_1\rho^{in}_2=0,Tr(\Pi_1\rho^{in}_1)>0$, and
 we have
\begin{align}
\begin{split}
&\Pi_1|\gamma_1\gamma_2\gamma_2\rangle=0,\Pi_1|\gamma_1\gamma_2\gamma_3\rangle=0,\Pi_1|\gamma_1\gamma_3\gamma_2\rangle=0,\\
&\Pi_1|\gamma_1\gamma_3\gamma_3\rangle=0,\Pi_1|\gamma_2\gamma_2\gamma_2\rangle=0,\Pi_1|\gamma_2\gamma_2\gamma_3\rangle=0,\\
&\Pi_1|\gamma_2\gamma_3\gamma_2\rangle=0,\Pi_1|\gamma_2\gamma_3\gamma_3\rangle=0,
\end{split}
 \end{align}
and
\begin{align}
&Tr(\Pi_1\rho^{in}_1)=\sum_{i,j,k=1}^{i,j,k=2}a_ib_ja_k\langle\gamma_i\gamma_{j+1}\gamma_k|\Pi_1|\gamma_i\gamma_{j+1}\gamma_k\rangle>0.
 \end{align}

(2) Let $\rho_1=\rho^{'}_2,\rho_2=\rho^{'}_1$. Then
$\rho^{in}_1=\rho^{'}_2\otimes \rho^{'}_1\otimes \rho^{'}_2,
\rho^{in}_2=\rho^{'}_2\otimes \rho^{'}_1\otimes \rho^{'}_1$.
According to (8), $\Pi_1\rho^{in}_2=0$, and we have
\begin{align}
\begin{split}
&\Pi_1|\gamma_2\gamma_1\gamma_1\rangle=0,\Pi_1|\gamma_2\gamma_1\gamma_2\rangle=0,\Pi_1|\gamma_2\gamma_2\gamma_1\rangle=0,\\
&\Pi_1|\gamma_2\gamma_2\gamma_2\rangle=0,\Pi_1|\gamma_3\gamma_1\gamma_1\rangle=0,\Pi_1|\gamma_3\gamma_1\gamma_2\rangle=0,\\
&\Pi_1|\gamma_3\gamma_2\gamma_1\rangle=0,\Pi_1|\gamma_3\gamma_2\gamma_2\rangle=0.
\end{split}
\end{align}

(3) Let $\rho_1=\rho^{'}_1,\rho_2=\rho^{'}_3$. Then
$\rho^{in}_1=\rho^{'}_1\otimes \rho^{'}_3\otimes \rho^{'}_1,
\rho^{in}_2=\rho^{'}_1\otimes \rho^{'}_3\otimes \rho^{'}_3$.
According to (8), $\Pi_1\rho^{in}_2=0$, and we have
\begin{align}
\begin{split}
&\Pi_1|\gamma_1\gamma_1\gamma_1\rangle=0,\Pi_1|\gamma_1\gamma_1\gamma_3\rangle=0,\Pi_1|\gamma_1\gamma_3\gamma_1\rangle=0,\\
&\Pi_1|\gamma_1\gamma_3\gamma_3\rangle=0,\Pi_1|\gamma_2\gamma_1\gamma_1\rangle=0,\Pi_1|\gamma_2\gamma_1\gamma_3\rangle=0,\\
&\Pi_1|\gamma_2\gamma_3\gamma_1\rangle=0,\Pi_1|\gamma_2\gamma_3\gamma_3\rangle=0.
\end{split}
\end{align}

(4) Let $\rho_1=\rho^{'}_3,\rho_2=\rho^{'}_1$, then
$\rho^{in}_1=\rho^{'}_3\otimes \rho^{'}_1\otimes \rho^{'}_3,
\rho^{in}_2=\rho^{'}_3\otimes \rho^{'}_1\otimes \rho^{'}_1$.
According to (8), $\Pi_1\rho^{in}_2=0$, and we have
\begin{align}
\begin{split}
&\Pi_1|\gamma_1\gamma_1\gamma_1\rangle=0,\Pi_1|\gamma_1\gamma_1\gamma_2\rangle=0,\Pi_1|\gamma_1\gamma_2\gamma_1\rangle=0,\\
&\Pi_1|\gamma_1\gamma_2\gamma_2\rangle=0,\Pi_1|\gamma_3\gamma_1\gamma_1\rangle=0,\Pi_1|\gamma_3\gamma_1\gamma_2\rangle=0,\\
&\Pi_1|\gamma_3\gamma_2\gamma_1\rangle=0,\Pi_1|\gamma_3\gamma_2\gamma_2\rangle=0.
\end{split}
\end{align}
Now using (10) and (12)-(14), we find that $Tr(\Pi_1\rho^{in}_1)$
in (11) is equal to zero, which contradicts (11) that is
$Tr(\Pi_1\rho^{in}_1)>0$. It means that there does not exist such
a fixed measurement. In other words, such an unambiguous
programmable  discriminator for mixed quantum states {\it does
not} exist. The proof is completed.

Why does not there exist such an unambiguous programmable
discriminator for mixed quantum states? The reason is not hard to
find from the above proof. It is because the mixed states
$\rho^{in}_1,\rho^{in}_2$ loose the symmetry which
$|\psi^{in}_1\rangle,|\psi^{in}_2\rangle$ have. Or we can say that
the difference between mixed states and pure states results in
\textit{Theorem 2}. Also, from \textit{Theorem
 2} we have seen some special features that mixed states have but
 pure states do not.

 \section{\label{Protocols1}Unambiguous discrimination between
mixed quantum states based on programmable discriminator}

It is disappointed that we do not have such an unambiguous
programmable discriminator for mixed quantum states that was
indicated above. We do not know whether there exists other type of
discriminators for mixed quantum states, either. However, if we
think about it from a different angle, we can find that the
unambiguous programmable discriminator is a very good idea for
discriminating states. We can still use the idea of unambiguous
programmable discriminators here to discriminate mixed states.
That is to say, if we want to discriminate two known mixed sates
$\rho_1,\rho_2$, then we can try to discriminate two mixed states
$\rho^{in}_1,\rho^{in}_2$. We use the idea of unambiguous
programmable discriminators which have two program registers and
$n$ data registers. Specifically, if we want to discriminate two
known mixed sates $\rho_1,\rho_2$, then we try to discriminate the
following states
\begin{align}
\begin{split}
&\rho^{in}_1=\rho_1\otimes \rho_2\otimes\rho^{\otimes
n}_1,\hskip1mm \rho^{in}_2=\rho_1\otimes
\rho_2\otimes\rho^{\otimes n}_2.
\end{split}
\end{align}
It is clear that if we can discriminate $\rho^{in}_1,\rho^{in}_2$,
then we can also discriminate $\rho_1,\rho_2$.

First we consider whether $\rho^{in}_1,\rho^{in}_2$ can be
unambiguously discriminated when $\rho_1,\rho_2$ can be
unambiguously discriminated. The answer is yes. We can use the
similar method in \textit{Theorem 1} to prove it. Now based on the
two known states $\rho^{in}_1,\rho^{in}_2$, we can construct POVMs
to distinguish them. Before dealing with the success probability
for unambiguous discrimination between $\rho^{in}_1$ and
$\rho^{in}_2$, we have a simple lemma as follows.

\textit{Lemma 1.} Let $\rho_1,\rho_2$ be two arbitrary mixed
states, and let $\rho^{in}_1=\rho_1\otimes
\rho_2\otimes\rho^{\otimes n}_1,\rho^{in}_2=\rho_1\otimes
\rho_2\otimes\rho^{\otimes n}_2$. We have
$F(\rho^{in}_1,\rho^{in}_2)=F(\rho_1,\rho_2)^{n}$, where $n\geq 1$
and $F(\cdot,\cdot)$ is the definition of fidelity in [1], i.e.,
$F(\rho_1, \rho_2)=Tr(\sqrt{\sqrt{\rho_1}\rho_2\sqrt{\rho_1}})$.

The proof of lemma 1 follows from the simple fact as follows.
\begin{align}
\begin{split}
&F(\rho_1\otimes\rho_2,\rho_3\otimes\rho_4)=F(\rho_1,\rho_3)\times
F(\rho_2,\rho_4).
\end{split}
\end{align}

Now we discuss the failure probability of the unambiguous
discrimination between $\rho^{in}_1,\rho^{in}_2$. According to
Raynal and L\"{u}tkenhaus' work [12], if $supp(\rho^{in}_1)\cap
supp(\rho^{in}_2)=\{0\}$ and some conditions are satisfied, the
failure probability of the unambiguous discrimination between
$\rho^{in}_1,\rho^{in}_2$ can reach its low bound. Let $F^{in}_1$
and $F^{in}_2$ denote
$\sqrt{\sqrt{\rho^{in}_1}\rho^{in}_2\sqrt{\rho^{in}_1}}$ and
$\sqrt{\sqrt{\rho^{in}_2}\rho^{in}_1\sqrt{\rho^{in}_2}}$,
respectively. Let $F(\rho^{in}_1,\rho^{in}_2)$ be the fidelity of
the two states $\rho^{in}_1,\rho^{in}_2$. Then
$F(\rho^{in}_1,\rho^{in}_2)=F(\rho_1,\rho_2)^n$. We denote by
$P^{in}_1$ and $P^{in}_2$, the projectors onto the supports of
$\rho^{in}_1$ and $\rho^{in}_2$, respectively. Let $P_1$ and $P_2$
be the projectors onto the supports of $\rho_1$ and $\rho_2$,
respectively. Then $P^{in}_1=P_1\otimes P_2\otimes P^{\otimes
n}_1$ and $P^{in}_2=P_1\otimes P_2\otimes P^{\otimes n}_2$. We can
prove $Tr(P^{in}_1\rho^{in}_2)=Tr(P_1\rho_2)^n$ and
$Tr(P^{in}_2\rho^{in}_1)=Tr(P_2\rho_1)^n$ using the similar method
as lemma 1. Let $\eta_1$ and $\eta_2$ be the priori probabilities
of $\rho_1$ and $\rho_2$, respectively. Now according to [12], we
have
\begin{widetext}
\begin{align}
\begin{split}
&Q^{opt}_{in}=\eta_2\frac{F(\rho_1,\rho_2)^{2n}}{Tr(P_2\rho_1)^n}
+\eta_1Tr(P_2\rho_1)^n \Leftrightarrow \left\{\begin{array}{r@{}l}
\rho^{in}_1-\frac{Tr(P_2\rho_1)^n}{F(\rho_1,
\rho_2)^n}F^{in}_1\geq0
\\ \rho^{in}_2-\frac{F(\rho_1,\rho_2)^n}{Tr(P_2\rho_1)^n}F^{in}_2\geq0 \end{array} \right. {\quad for
\quad} \sqrt{\frac{\eta_2}{\eta_1}}\leq\frac{Tr(P_2\rho_1)^n}{F(\rho_1,\rho_2)^n}, \\
&Q^{opt}_{in}=2\sqrt{\eta_1\eta_2}F(\rho_1,\rho_2)^{n}
\Leftrightarrow \left\{\begin{array}{r@{}l}
\rho^{in}_1-\sqrt{\frac{\eta_2}{\eta_1}}F^{in}_1\geq0
\\ \rho^{in}_2-\sqrt{\frac{\eta_1}{\eta_2}}F^{in}_2\geq0 \end{array} \right.
{\quad for \quad}
\frac{Tr(P_2\rho_1)^n}{F(\rho_1,\rho_2)^n}\leq\sqrt{\frac{\eta_2}{\eta_1}}\leq
\frac{F(\rho_1,\rho_2)^n}{Tr(P_1\rho_2)^n},\\
&Q^{opt}_{in}=\eta_1\frac{F(\rho_1,\rho_2)^{2n}}{Tr(P_1\rho_2)^n}
+\eta_2Tr(P_1\rho_2)^n\Leftrightarrow \left\{\begin{array}{r@{}l}
\rho^{in}_1-\frac{F(\rho_1,\rho_2)^n}{Tr(P_1\rho_2)^n}F^{in}_1\geq0
\\ \rho^{in}_2-\frac{Tr(P_1\rho_2)^n}{F(\rho_1,\rho_2)^n}F^{in}_2\geq0 \end{array} \right. {\quad for \quad}
\frac{F(\rho_1,\rho_2)^n}{Tr(P_1\rho_2)^n}\leq\sqrt{\frac{\eta_2}{\eta_1}},
\end{split}
\end{align}
\end{widetext}
where $Q^{opt}_{in}$ denotes the optimal failure probability of
the unambiguous discrimination between $\rho^{in}_1,\rho^{in}_2$.
Here $Tr(P_2\rho_1)\leq 1$, $Tr(P_1\rho_2)\leq 1$,
$F(\rho_1,\rho_2)^2\leq Tr(P_2\rho_1)$ and $F(\rho_1,\rho_2)^2\leq
Tr(P_1\rho_2)$ (the more details are referred to [12]).

The first question is whether  or not $supp(\rho^{in}_1)\cap
supp(\rho^{in}_2)=\{0\}$ can be satisfied? Actually, we can easily
prove that if $supp(\rho_1)\cap supp(\rho_2)=\{0\}$, then
$supp(\rho^{in}_1)\cap supp(\rho^{in}_2)=\{0\}$. It means that
$supp(\rho^{in}_1)\cap supp(\rho^{in}_2)=\{0\}$ is not a stricter
constraint.

Let $Q_{in}$ denote the failure probability of the unambiguous
discrimination between $\rho^{in}_1,\rho^{in}_2$. From [12], we
know that  $Q_{in}$ here can reach $Q^{opt}_{in}$ sometimes. When
$Q_{in}$ reaches $Q^{opt}_{in}$, that is, $Q_{in}=Q^{opt}_{in}$,
we find that $Q_{in}$ is smaller than $Q$ (here $Q$ denotes the
failure probability of the unambiguous discrimination between
$\rho_1,\rho_2$), except for some special cases. We discuss this
 in what follows.

If $F(\rho_1,\rho_2)=0$, i.e., it means that the two states can be
perfectly discriminated, then $Q_{in}=Q=0$. When $n=1$, we find
that if $Q_{in}$ reaches $Q^{opt}_{in}$, then $Q$ can also reach
its optimal value, and thus $Q_{in}=Q=Q^{opt}_{in}$. Now we
consider the situation where $0<F(\rho_1,\rho_2)<1$ and $n>1$:

(1) If $\frac{Tr(P_2\rho_1)}{F(\rho_1,\rho_2)}\leq 1$ and $
\frac{F(\rho_1,\rho_2)}{Tr(P_1\rho_2)}\geq 1$, then no matter
which regime $\sqrt{\frac{\eta_2}{\eta_1}}$ is, we will find that
if $Q_{in}$ reaches $Q^{opt}_{in}$, then $Q_{in}=Q^{opt}_{in}<Q$.

(2) If
$\frac{Tr(P_2\rho_1)}{F(\rho_1,\rho_2)}\leq\frac{F(\rho_1,\rho_2)}{Tr(P_1\rho_2)}<
1$, then, except for the regime
$\frac{F(\rho_1,\rho_2)^n}{Tr(P_1\rho_2)^n}\leq\sqrt{\frac{\eta_2}{\eta_1}}\leq\frac{F(\rho_1,\rho_2)}{Tr(P_1\rho_2)}$
that we cannot compare, we will find that if $Q_{in}$ reaches
$Q^{opt}_{in}$, then $Q_{in}=Q^{opt}_{in}<Q$.

(3) If $\frac{Tr(P_2\rho_1)}{F(\rho_1,\rho_2)}>1$, then, except
for the regime
$\frac{Tr(P_2\rho_1)}{F(\rho_1,\rho_2)}\leq\sqrt{\frac{\eta_2}{\eta_1}}\leq\frac{Tr(P_2\rho_1)^n}{F(\rho_1,\rho_2)^n}$
that we cannot compare, we will find that if $Q_{in}$ reaches
$Q^{opt}_{in}$, then $Q_{in}=Q^{opt}_{in}<Q$.

From the above discussion we can see that if the failure
probability of the unambiguous discrimination between
$\rho^{in}_1,\rho^{in}_2$ reaches its optimization, then the
failure probability of the unambiguous discrimination between
$\rho^{in}_1,\rho^{in}_2$ is better than that between
$\rho_1,\rho_2$ mostly. It is easy to find that the bigger $n$ is,
the smaller $Q^{opt}_{in}$ will be. That means that if $Q_{in}$
can reach $Q^{opt}_{in}$ with the bigger $n$, then the smaller
$Q_{in}$ will be. Considering the conditions of $Q_{in}$ being
able to reach $Q^{opt}_{in}$ in (17), we find that such conditions
are not stricter when $n$ is bigger. Especially, the conditions in
the first and the third regime of (17) can be derived from $n=1$.
On the other hand, even if $n$ is small, such as $n=2$, and
$F(\rho_1,\rho_2)$ is much smaller than 1, then we can also have a
very small $Q^{opt}_{in}$ here.

A rest question is what about the situation when $Q_{in}$ does not
reach its optimization? We have no answer yet. The solution  of
such a question depends on the solution of how to discriminate two
arbitrary mixed states optimally. However, how to discriminate
optimally two arbitrary mixed quantum states  still is  an open
question now.

 \section{conclusions}
 In this paper, we try to design an unambiguous programmable discriminator for mixed quantum states based on Bergous and
Hillery's  model [15]. We have proved that there does not exist a
universal unambiguous programmable discriminator for mixed quantum
states that has two program registers and one data register.
However, we found that we can use the idea of programmable
discriminators to unambiguously discriminate mixed quantum states.
The research shows that by using such an idea, when the  success
probability for discrimination reaches  the upper bound, the
success probability is better than what we do not use the idea to
do, except for some special cases. We have discussed this result
in detail and presented some prospects for it.

 This work is
supported in part by the National Natural Science Foundation (Nos.
90303024, 60573006), the Higher School Doctoral Subject Foundation
of Ministry of Education (No. 20050558015), and the Natural
Science Foundation of Guangdong Province (No. 031541) of China.

\end{document}